# Single photonic perceptron based on a soliton crystal Kerr microcomb for high-speed, scalable, optical neural networks

Xingyuan Xu,[1] Mengxi Tan,[1] Bill Corcoran,[2] Jiayang Wu,[1] Thach G. Nguyen,[3] Andreas Boes,[3] Sai T. Chu,[4] Brent E. Little,[5] Roberto Morandotti,[6] Arnan Mitchell,[3] Damien G. Hicks,[1,7] and David J. Moss[1,]*

**Affiliations:**

[1]Optical Sciences Centre for Micro-Photonics, Swinburne University of Technology, Hawthorn, VIC 3122, Australia

[2]Department of Electrical and Computer Systems Engineering, Monash University, Clayton, 3800 VIC, Australia

[3]School of Engineering, RMIT University, Melbourne, VIC 3001, Australia

[4]Department of Physics and Material Science, City University of Hong Kong, Tat Chee Avenue, Hong Kong, China.

[5]Xi'an Institute of Optics and Precision Mechanics Precision Mechanics of CAS, Xi'an, China.

[6]INRS-Énergie, Matériaux et Télécommunications, 1650 Boulevard Lionel-Boulet, Varennes, Québec, J3X 1S2, Canada.

[7]Bioinformatics Division, Walter & Eliza Hall Institute of Medical Research, Parkville, Victoria 3052, Australia

*Correspondence to:  dmoss@swin.edu.au

Abstract

Optical artificial neural networks (ONNs) — analog computing hardware tailored for machine learning [1, 2] — have significant potential for ultra-high computing speed and energy efficiency [3]. We propose a new approach to architectures for ONNs based on integrated Kerr micro-comb sources [4] that is programmable, highly scalable and capable of reaching ultra-high speeds. We experimentally demonstrate the building block of the ONN — a single neuron perceptron — by mapping synapses onto 49 wavelengths of a micro-comb to achieve a high single-unit throughput of 11.9 Giga-FLOPS at 8 bits per FLOP, corresponding to 95.2 Gbps. We test the perceptron on simple standard benchmark datasets — handwritten-digit recognition and cancer-cell detection — achieving over 90% and 85% accuracy, respectively. This performance is a direct result of the record small wavelength spacing (49GHz) for a coherent integrated microcomb source, which results in an unprecedented number of wavelengths for neuromorphic optics. Finally, we propose an approach to scaling the perceptron to a deep learning network using the same single micro-comb device and standard off-the-shelf telecommunications technology, for high-throughput operation involving full matrix multiplication for applications such as real-time massive data processing for unmanned vehicle and aircraft tracking.

# Introduction

Artificial Neural Networks (ANNs) have demonstrated unprecedented success in making predictions from, and capturing simpler representations of, complex high-dimensional data. When trained on enough data, ANNs can outperform humans and other computational algorithms [5] in tasks ranging from image recognition and language translation to risk evaluation and, interestingly, sophisticated board games [6]. The computing power and speed of ANNs is dictated by matrix multiplication operations. Current electronic devices designed for ANNs, including the IBM TrueNorth and Google TPU [7, 8], generally employ ultra-large-scale parallel arrays of processors, such as the systolic array [8], to enhance the parallelism for higher computing speeds. Electronic approaches, however, are subject to either relatively inefficient digital protocols or the electrical bandwidth bottleneck of each single processor (~700 MHz) [9].

Photonic ANN hardware, or optical neural networks (ONNs), are promising next-generation neuromorphic processors, since they potentially offer ultra-large optical bandwidths in order to reach dramatically accelerated computing speeds [3]. The key to achieving ONNs is to realize the weighted synapses that connect the neurons and nodes. Unlike digital approaches where the synapses are stored in memories, photonic approaches not only rely on physical embodiments of synapses where the number of synapses (i.e., the network scale) relies on the physical parallelism, but it is inherently analog.

Significant progress has been made on ONNs that explore different multiplexing approaches to realise parallel synapses. Spatially multiplexed schemes such as integrated coherent photonic circuits [3] and diffractive frameworks [10], have successfully demonstrated classification tasks involving vowels and handwritten digits with low-power passive operation, although with a direct tradeoff between parallelism and footprint. Other approaches to ONNs, such as photonic reservoir computing [11-13] and spike processing [14-17], employ advanced multiplexing techniques to establish synapses with much more compact schemes. Photonic reservoir computing begins with using time-domain multiplexing to achieve large-scale input layers with hundreds of nodes. Spike processing, on the other hand, employs wavelength-division multiplexing and has achieved significant success at pattern recognition tasks through the use of integrated phase-change devices [17].

Despite these successes, however, current schemes have limitations of one form or another. Time-division multiplexed networks are difficult to either dynamically train or scale up to form deep (multi-layer) neural networks. Spike processing has so far been limited in its degree of parallelism by the use of discrete laser arrays. The simultaneous use of all three types of multiplexing (wavelength, time, spatial) would offer the greatest benefits in terms of scale and processing power and speed.

Here, we propose a new approach to ONNs, based on a Kerr frequency optical micro-comb source in an integrated micro-ring resonator (MRR). This approach to ONNs employs the combined use of wavelength, time and spatial multiplexing, and can perform matrix multiplication operations at high throughput speed and in an intrinsically scalable and dynamically trainable network structure. We experimentally demonstrate the key building block of the full ONN, a single neuron photonic perceptron with 49 synapses, that operates at a single-unit matrix multiplication (vector dot product) throughput of 11.9 Giga floating point operations/s (FLOPS), that, at 8 bits per FLOP, corresponds to a bit rate of 95.2 Gbps. We achieve this record throughput for ONNs by simultaneously weighting the synapses in the wavelength domain and scaling the input data in the time domain, enabled by the microcomb source. We apply the single perceptron to standard benchmark tests including the classification of handwritten digits, where we achieve > 93% accuracy, as well as to predicting benign/malignant cancer classes using a feature set extracted from microscopy images of biopsied tissue, achieving > 85% accuracy.

Finally, we show how this approach can be readily scaled using the same single micro-comb chip source to form ultrahigh speed deep neural networks using standard off-the-shelf telecommunications tools. Scaling to multiple levels offers the full potential of wavelength multiplexing for speed enhancement together with the deep learning network structure. Both the perceptron and deep learning ONN are dynamically trainable and fully compatible with state-of-art electrical interfaces, making them highly promising for next generation real-time massive data processing.

## Photonic single perceptron

Figure 1 shows the mathematical model of the single neuron perceptron [18] while Figure 2 shows the detailed experimental configuration that we use based on an integrated optical micro-comb source. The perceptron uses simultaneous time and wavelength multiplexing based on 49 wavelengths from the microcomb source, each wavelength forming a single synapse. Its core function is a matrix multiplication operation (for a single perceptron, reducing to a vector dot-product) between the input electronic data of the image to be analysed with the synaptic weights that are implemented in a multiple-step approach in the optical domain. The raw input data for classification is a 28×28 matrix in electronic digital grey-scale values with 8-bit intensity resolution. We first resample this digitally (effectively performing digital down-sampling) into a 7×7 matrix which is then rearranged into a 1D vector: $\mathbf{X} = [x(1), x(2), \ldots, x(49)]$. This vector is then sequentially multiplexed in the time-domain via a high-speed electrical digital-to-analog converter at a data rate of 11.9 Giga Baud (symbols per second), where each symbol corresponds to the 8-bit pixel of the input data and occupies one time-slot of length $\tau = 84$ ps, so that the entire waveform duration is given by $N\tau = 4.12$ ns ($N=49$). In traditional digital approaches, the input nodes to the neural network generally reside in electronic memories and are routed via the memory addresses. In contrast, for our ONN the input nodes are defined by temporally multiplexed symbols that can be routed according to their temporal location.

Next, the electronic time-division multiplexed input waveform signal is multicast onto all 49 (e.g. equal to the number of components of the X vector) wavelength channels from the micro-comb source via an electro-optic modulator, such that each wavelength contains an identical replica of the temporal data waveform **X**. The optical power of each comb line is then weighted with an optical spectral shaper (Waveshaper) according to the trained synaptic weight vector $\mathbf{W} = [w(1), w(2), \ldots, w(49)]$, which therefore effectively multiplexes the synaptic weights in the wavelength domain. Assuming **X** and **W** are both 49×1 column vectors, the resulting weighted replicas of input **X** then become

$$\mathbf{X} \times \mathbf{W}^T = \begin{pmatrix} w(1) \cdot x(1) & w(1) \cdot x(2) & w(1) \cdot x(3) & \cdots & w(1) \cdot x(49) \\ w(2) \cdot x(1) & w(2) \cdot x(2) & w(2) \cdot x(3) & \cdots & w(2) \cdot x(49) \\ w(3) \cdot x(1) & w(3) \cdot x(2) & w(3) \cdot x(3) & \cdots & w(3) \cdot x(49) \\ \vdots & \vdots & \vdots & \ddots & \vdots \\ w(49) \cdot x(1) & w(49) \cdot x(2) & w(49) \cdot x(3) & \cdots & w(49) \cdot x(49) \end{pmatrix} \quad (1)$$

where the $n_{\text{th}}$ row ($n \in [1, N]$) corresponds to the weighted temporal waveform replica at the $n_{\text{th}}$ wavelength channel. Hence, the diagonal elements denote the $N$ weighted input nodes, i.e., the $n_{\text{th}}$ weighted input node is represented by the 8-bit symbol $w(n) \cdot x(n)$ residing in the $n_{\text{th}}$ timeslot of the $n_{\text{th}}$ wavelength channel.

The replicas then pass through a dispersive element providing second-order dispersion to progressively delay the weighted replicas so as to line up all of the diagonal elements into the same timeslot, with the delay step satisfying $\tau = \text{delay}(\lambda_k) - \text{delay}(\lambda_{k+1})$. Thus, the dispersive element serves as a time-of-flight addressable memory that aligns the sequentially weighted temporal symbols $w(1) \cdot x(1), w(2) \cdot x(2), \ldots, w(49) \cdot x(49)$ across the wavelength channels as

$$\begin{pmatrix} & & & & w(1)\cdot x(1) & \cdots & w(1)\cdot x(47) & w(1)\cdot x(48) & w(1)\cdot x(49) \\ & & & w(2)\cdot x(1) & w(2)\cdot x(2) & \cdots & w(2)\cdot x(48) & w(2)\cdot x(49) \\ & & w(3)\cdot x(1) & w(3)\cdot x(2) & w(3)\cdot x(3) & \cdots & w(3)\cdot x(49) \\ & \iddots & \vdots & \vdots & \vdots & \iddots \\ w(49)\cdot x(1) & \cdots & w(49)\cdot x(47) & w(49)\cdot x(48) & w(49)\cdot x(49) \end{pmatrix}$$

(2)

While for the single perceptron demonstrated here (single layer, single neuron) this process does not increase the speed of the network since only the diagonal elements are used, dramatic increases in speed can be realized by scaling to deep networks through simultaneous time, wavelength and spatial multiplexing (see section 5).

Finally, the optical intensity of the aligned time slots are summed by photodetection (high speed – with enough bandwidth to resolve the different timeslots of width τ) and sampling to finally yield the result for matrix multiplication (in this case a vector dot product) of the neuron, given by

$$\mathbf{X}\cdot\mathbf{W} = \sum_{k=1}^{49} w(k)\cdot x(k) \qquad (3)$$

After this matrix multiplication, the weighted and summed output is then biased and mapped onto a desired range through a nonlinear sigmoid function (achieved in this initial demonstration offline with digital electronics, see supplementary for details), yielding the neuron (single-neuron perceptron) output. Finally, the prediction of the input data's category is generated by comparing the neuron output with the decision boundary, which is a hyper-plane in a 49-dimension space found during the learning process (in this case achieved offline digitally) that can well separate the two input categories.

## Soliton crystal microcomb

The key to our approach lies in the use of an integrated optical micro-comb source [19-21]. Micro-combs have enabled many fundamental breakthroughs through their ability to generate optical signals with the same precision as microwave and RF signals, yet at 100's of terahertz for optical frequency synthesis [22], ultrahigh capacity communications [23], complex quantum state generation [24], advanced microwave signal processing [25], and more. They offer the full power of optical frequency combs [26] but in an integrated platform with much smaller footprint and higher scalability, performance, and reliability [27-35].

The microcomb we employ here operates in a unique coherent state termed "soliton crystals", which originate from optical parametric oscillation in an on-chip micro-ring resonator (MRR). Soliton crystals are a unique and powerful class of soliton microcomb featuring deterministic formation originating from a mode crossing-induced background wave, driven by the Kerr nonlinearity, together with the high intra-cavity power. Because the intracavity energy of the soliton crystal state is almost identical to that of the chaotic state from which they originate, there is no significant change in intracavity energy when they are generated and, in turn, there is no resulting self-induced shift that requires complex tuning methods as, e.g., for DKS solitons [27]. This results in simple and reliable initiation via adiabatic pump wavelength sweeping [36], as well as much higher energy efficiency (ratio of optical power in the comb lines relative to the pump power) [37]. Soliton crystals are thus a very promising category of optical frequency combs for wavelength multiplexing based systems including microwave and RF photonic processors [25, 38-53] as well as the ONN reported here.

The MRR used here was fabricated in a CMOS compatible doped silica glass platform [20] with a Q factor of ~1.5 million and radius of ~592 μm, corresponding to an FSR of ~0.4 nm or 48.9 GHz. This is a record low FSR spacing for any coherent integrated microcomb source and is a critical feature of this work since it resulted in a large number of available wavelengths over the telecommunications C-band. The chip was coupled with a fibre array, featuring a fibre-chip coupling loss of only 0.5 dB per facet brought about by integrated mode converters. The cross-section of the waveguide was designed to be 3 μm × 2 μm, which yielded anomalous dispersion in the C band as well as a unique mode crossing observed at ~ 1552 nm.

To generate coherent micro-combs, a CW pump laser was employed, with the power amplified to 30dBm by an optical amplifier. Next, the wavelength was subsequently manually swept from blue to red. When the detuning between pump wavelength and MRR's cold resonance became small enough such that the intra-cavity field reached a threshold value, a modulation instability driven oscillation was initiated. As the detuning was changed further, distinctive 'fingerprint' optical spectra (Fig. 3) were observed that are a signature of soliton crystals [36, 37], which arise from spectral interference between the tightly packed solitons circulating along the ring cavity.

## Experimental results

We experimentally demonstrated the building block of the network – a single layer, single neuron photonic perceptron (Fig. 4) which is suitable for binary classification problems. Problems with more classes can be addressed using more than one neuron, even with only a single layer (non-deep) ONN system. This can easily be achieved by sub-dividing the comb into wavelength groups that each define a neuron (See Section 5). We first tested the perceptron on several pairs of handwritten digits (Fig. 5 and 6), using 500 figures for each digit, from which 920 figures were randomly selected for offline pre-training, leaving the remaining 80 figures for experimental testing. The 2D handwritten digit figures were pre-processed electronically using a down-sampling method to reduce the image size from 28×28 to 7×7, followed by transforming it into a one-dimensional array of 49 symbols. This was then time multiplexed with ~ 84 ps long timeslots for each symbol (Fig. 5b), equating to a modulation speed of 11.9 Giga-baud.

As discussed above, the optical power of the 49 microcomb lines was shaped according to pre-learned synaptic weights (Fig. 6a) to boost the parallelism and establish synapses for the neuron. Then the input data stream was multicast onto all 49 shaped comb lines followed by a progressive (linear with wavelength) delay using a ~13 km standard single-mode fibre (SMF), which served as the time-of-flight optical buffer via its second-order dispersion (~17 ps/nm/km). Hence, the weighted symbols on different wavelength channels were aligned temporally, allowing them to be summed together via photodetection and sampling of the central timeslot, to generate the results of the matrix multiplication and accumulate (MAC) operation. The output was then compared with the decision boundary obtained from the learning process, which yielded the final ONN prediction (Fig. 6b).

We evaluated the performance of the optical perceptron in determining the classification of two standard benchmark cases (see Supplementary and Figures 6, 7), handwritten digits and cancer cells. In the first case, two categories of handwritten digits (0 and 6) were distinguished by the decision boundary. Our device achieved an accuracy (ACC) of 93.75%, compared to 98.75% success for the calculated results on a digital computer (see Fig. 6d). Despite being a rudimentary benchmark tests, the perceptron nevertheless achieved a very high success rate and, most importantly, at unprecedented speeds (see below). This was a result of the large number of synapses (optical wavelengths over the C-band), in turn enabled by the record low FSR soliton crystal micro-comb.

We also determined the classification of cancer cells from tissue biopsy data (Fig. 6e and supplementary). Individual cell nuclei, from breast mass tissue extracted by fine needle aspirate and imaged under a microscope, have previously been characterized in terms of 30 features such as radius, texture, perimeter, etc. In our analysis, data for 521 cell nuclei were employed for pre-training, with another 75 used for experimental diagnosis, following a similar procedure to the above handwritten digit test. We achieved an accuracy of 86.67% as compared to 98.67% success for the calculated results on a digital computer.

There is currently no commonly accepted standard that establishes benchmark systems for classifying and quantifying the computing speed and processing power of the widely varying types of ONNs that have been reported. Therefore, we explicitly outline the performance definitions that we use for throughput and latency (see Supplementary) in characterizing our ONN. We follow the approach Intel has used to evaluate digital micro-processors [54]. Considering that in our system the input data and weight vectors for the MAC calculation originate from different paths and are interleaved in different dimensions (time, wavelength), we use the temporal sequence at the electrical output port to clearly define the throughput. According to the broadcast-and-delay protocol, each computing cycle of matrix multiplication between the 49-symbol data and weight vectors generates an output temporal sequence with a length of 48+1+48 symbols and thus a total time duration of 84ps×97. While the 49$^{th}$ symbol corresponds to the desired matrix multiplication output as a result of 49 multiply-and-accumulate operations, the throughput of our ONN is thus given as (49×2)/(84 ps×97)=11.9 Giga-FLOPS.

The input data stream consisted of symbols with 8-bit (256 discrete levels) values, determined by both the original grey scale values of the image pixels and the intensity resolution of our electronic arbitrary waveform generator. The optical spectral shaper (Waveshaper) featured an attenuation control range of 35 dB, which could support up to 11-bit resolution (10·log10($2^{11}$)=33 dB). As such, each computing cycle also corresponded to an equivalent throughput of (49×2)×8/(84 ps×97)=95.2 Gbps in terms of bit rate. For analog systems such as the one used here, the bit rate/intensity resolution is limited by the signal-to-noise ratio of the system. Hence, to achieve 8-bit resolution, the system would have to feature a signal-to-noise ratio of >20·log10($2^8$)=48 dB in electrical power or 24 dB in optical power. This is well within the capability of analog microwave photonic links including the ONN system reported here (with OSNR >28 dB).

Our results represent the fastest throughput (in bit rate) claimed so far for any ONN, although a direct comparison of the widely varying systems is difficult (Supplementary Table S1). For example, while systems that use CW sources to perform single-shot measurements [4, 10, 17] may have a low *latency*, they always suffer from a very low *throughput* since the input data cannot be updated rapidly. While the *latency* of our single perceptron is relatively high (~64 μs) due to the fibre spool, this does not affect the *throughput* of our system. In any event the latency can be readily reduced to < 200 ps through the use of compact devices to implement the delay function — devices with high group velocity dispersion and much lower overall time delay such as photonic crystal waveguides or sampled Bragg gratings (in fibre or on-chip) [55], for example. Finally, although we implemented the nonlinear function digitally offline, which did not impact the predictions, this could also be done with electro-optical modulators or electrical amplifiers operating at saturation point.

## Scaling to deep ONNs

The single neuron perceptron can be readily scaled, using many different approaches, to multi-layer deep ONNs using only the same single micro-comb source together with standard off-the-shelf telecommunications technologies. Deep neural networks can achieve much more complex tasks than the single perceptron demonstrated here, and at much higher speeds. Here, we outline in detail one possible example of a scaled deep learning network (Fig. 7). It consists of an input layer (serving as an interface between the input raw data and the neural network), multiple hidden layers (each containing multiple neurons) and an output layer. The deep ONN also uses wavelength division multiplexing to establish the synapses but, in contrast with the single perceptron, makes full use of time, wavelength and spatial multiplexing with all layers' synapses being established from the same single soliton crystal source and using the same single WaveShaper device. The microcomb is replicated and spatially multiplexed into the multiple hidden layers with each layer (and each synapse, or wavelength, within each layer) all being uniquely weighted. The simultaneous power splitting and spectral shaping can be achieved with a single commercially available Waveshaper. At each layer, the comb is further divided spectrally with into $M(k)$ groups, where $M(k)$ is the number of neurons and $k$ is the layer number, with each group defining one neuron. Because the neurons are defined by their wavelength sub-comb, rather than physically, in effect they are "virtual"— as are the synapses. The layers each have an electrical input port to receive the electrical output of the previous layer and an electrical output port to generate the calculated results of the current layer. The network is scalable in that each hidden layer can have a different number of neurons and synapses. The only requirement, assuming a fully connected network, is that the number of synapses for each neuron — equal to the number of neurons in the previous layer $M(k-1)$ — needs to satisfy the relation $M(k-1) \cdot M(k) \leq N_{comb}$ where $N_{comb}$ is the total number of generated microcomb lines.

Figure 7b shows what the signal looks like at the different locations in the network. The weighted combs (neuron synapses) input to each layer are modulated by the time multiplexed electrical signal (with a waveform duration of $T_{in}= M(k-1) \cdot \tau$, where $\tau$ is each symbol's duration) input from the previous layer. Following this, the WDM waveforms for each neuron are progressively delayed by a dispersive device. In contrast with the single perceptron where all time slots in the full comb need to be aligned to a single slot (since there is only one neuron), here only the wavelengths within each individual neuron need to be aligned. One of the most elegant methods to achieve this would be through the use of chirped sampled Bragg gratings. Each segment of the grating individually serves as the buffer for the wavelengths associated with each neuron, with the delay between wavelengths matching the symbol duration of the input electrical waveform — both equal to $\tau$. The sampled Bragg grating is not only capable of imposing segmented delays on many wavelengths simultaneously, but does so without any significant overall delay, or *latency*. The delayed replicas of each neuron are then demultiplexed in wavelength and summed separately via photo-detection. Since the network uses spatial multiplexing to address the different hidden layers, it requires multiple delay components (e.g., chirped sampled Bragg gratings)—equal to the number of hidden layers. We note that since the different layers can have different numbers of neurons, the grating structure would also be layer dependent — the number of segments must equal the number of neurons while the bandwidth of each segment depends on the number of synapses.

The last stage consists of digital signal processing the electrical waveforms generated by the previous levels of neurons to sample the central summing slot to obtain the matrix multiplication result of each neuron. This is followed by imposing a nonlinear function that rescales the weighted sum, and finally by retiming and digital-to-analog (D/A) conversion to generate the final output of the layer, with a time duration of $T_{out}= M(k) \cdot \tau$ and a modulation rate equal to the electrical input waveform. Note that while the digital signal processing adds to the overall *latency*, it does not affect the net *throughput* rate. The resampling needed to preserve the input data rate of each layer can easily be achieved with high-speed electronic circuits (such as field programmable gate arrays (FPGAs)) or potentially even using optical approaches [56].

After the sequence processing by the different layers, the ONN then predicts the class of the raw input data as before by comparing with multi-dimensional hyperspace decision boundaries determined through prior training. This overall network structure results in a series of wavelength, time and spatially multiplexed signals that dramatically boosts the network scale to multiple hidden layers each having multiple neurons, operating at ultra-high speed, and yet within a compact footprint. The potential throughput of the deep ONN can easily reach the TeraFLOP/s regime, and be capable of solving much more complex tasks than the ones achieved by the single perceptron demonstrated here — see Supplementary section for detail.

Finally, there is strong potential for substantially higher levels of integration — ultimately towards fully monolithic embodiments of our ONN. The central component of our system, the optical frequency comb source, is already integrated, while all of the other components have been demonstrated in integrated forms, including integrated InP spectral shapers [57], high-speed integrated lithium niobate modulators [58], integrated dispersive elements [59], and photodetectors [60]. Finally, low power-consumption (98mW) Kerr combs have recently been demonstrated [61], that would greatly reduce the energy requirements.

## Conclusion

We propose a novel and powerful approach to optical neural networks based on integrated optical Kerr micro-comb sources. We demonstrate the key building block - a single layer, single neuron perceptron - operating at a record single-unit throughput of 11.9 GFLOPS or 95.2 Gbps. We successfully perform standard benchmark real-life tasks including the recognition of handwritten digits and the diagnosis of cancer cells. We propose a specific architecture to realize a deep learning ONN with greatly enhanced throughput speed and processing power, enabled by the high degree of parallelism achieved through simultaneous wavelength, time, and spatial multiplexing. This approach offers significant potential for real-time analysis of high-dimensional data, such as identifying astrophysical fast radio bursts or pathology identification in clinical scanning applications [62, 63].

**Acknowledgments:** This work was supported by the Australian Research Council Discovery Projects Program (No. DP150104327). R. M. acknowledges support by the Natural Sciences and Engineering Research Council of Canada (NSERC) through the Strategic, Discovery and Acceleration Grants Schemes, by the MESI PSR-SIIRI Initiative in Quebec, and by the Canada Research Chair Program. Brent E. Little was supported by the Strategic Priority Research Program of the Chinese Academy of Sciences, Grant No. XDB24030000. D.G.H was supported in part by Australian Research Council grant FT104101104.

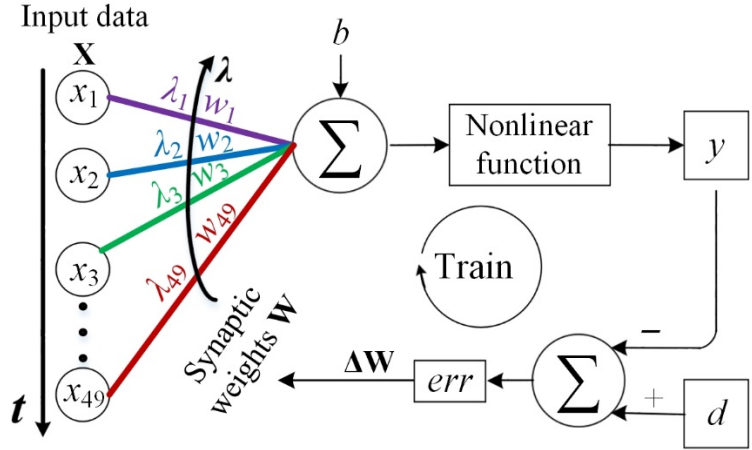

Fig. 1. Mathematical model of the perceptron. The perceptron featured 49 input nodes X = [$x(1)$, $x(2)$, …, $x(49)$], which were connected to the neuron with 49 reconfigurable weights W = [$w(1)$, $w(2)$, …, $w(49)$]. After the matrix multiplication, the input data X was weighted and summed, then added with a bias $b$ and passed through a nonlinear sigmoid function to generate the output $y$. The output $y$ was compared with the desired output $d$ to generate an error signal $err$ to adjust the weights. The training was performed offline.

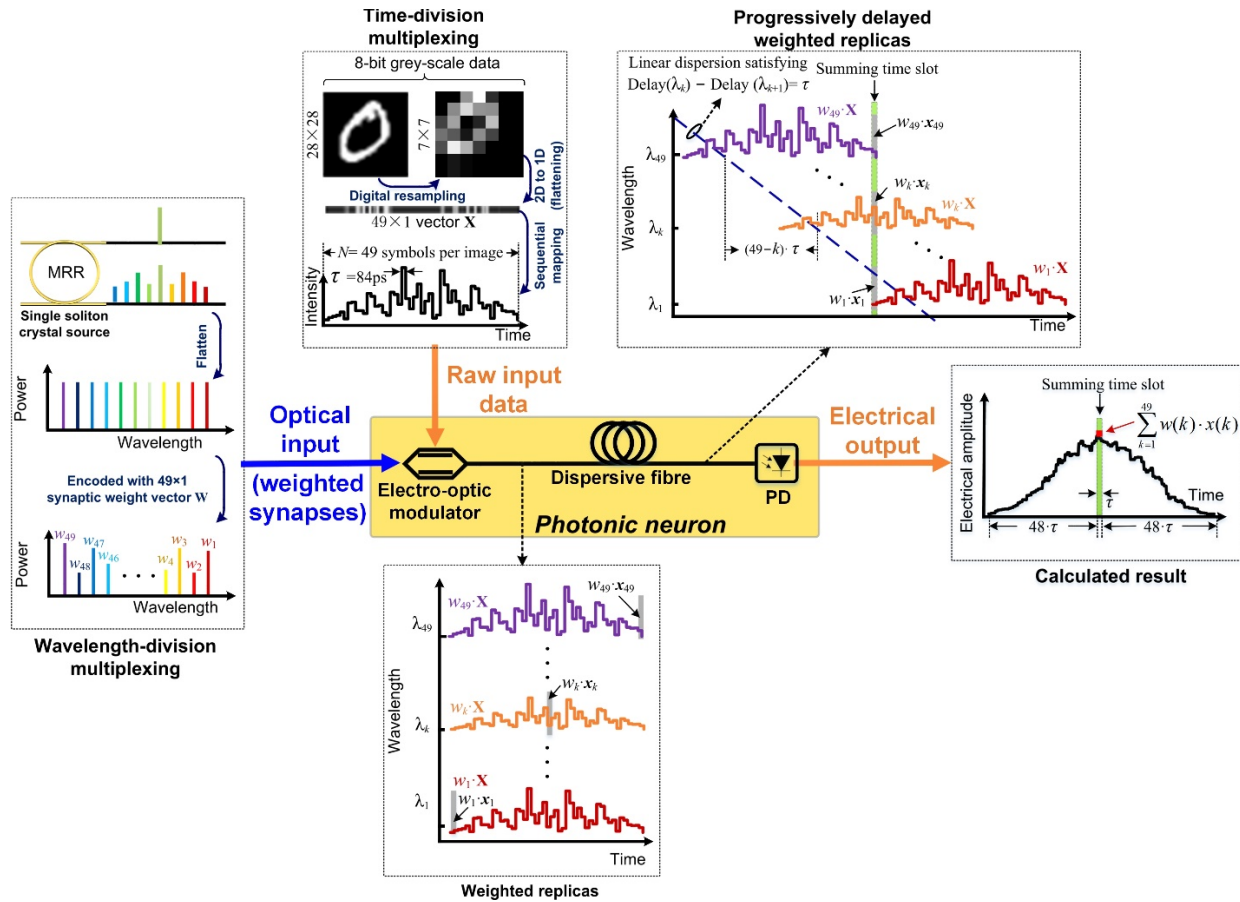

Fig. 2. Operation principle of the perceptron or single photonic neuron. PD: photodetector.

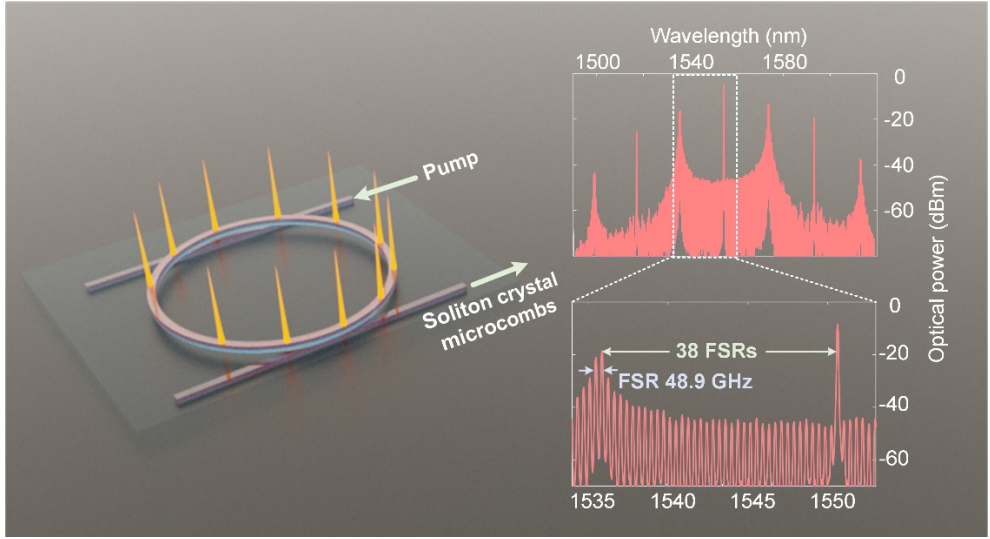

Fig. 3. Schematic of soliton crystal microcombs and generated optical spectrum. The soliton crystal is generated in a 4-port integrated micro-ring resonator (MRR) with an FSR of 49GHz.

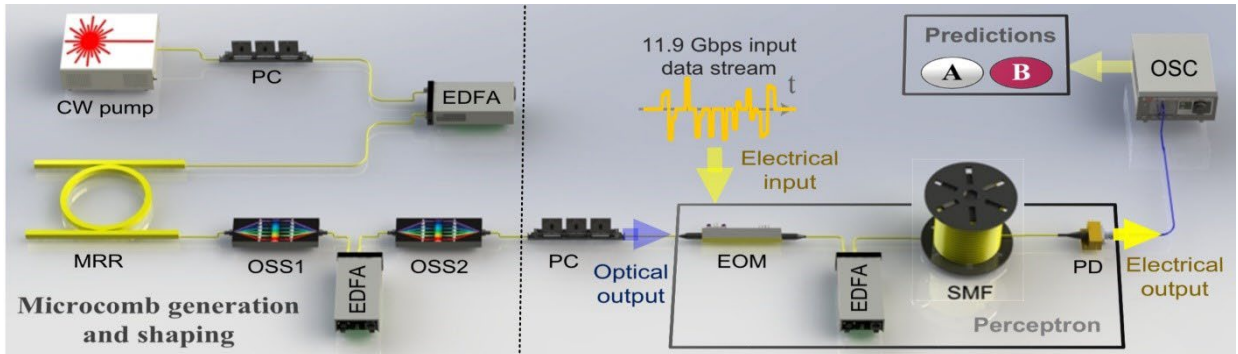

Fig. 4. The experiment setup of the optical perceptron. CW pump: continues-wave pump laser. PC: polarization controller. EDFA: erbium doped fibre amplifier. MRR: micro-ring resonator. OSS: optical spectral shaper. EOM: electro-optical Mach-Zehnder modulator. SMF: standard single mode fibre for telecommunications. PD: photodetector. OSC: high-speed oscilloscope.

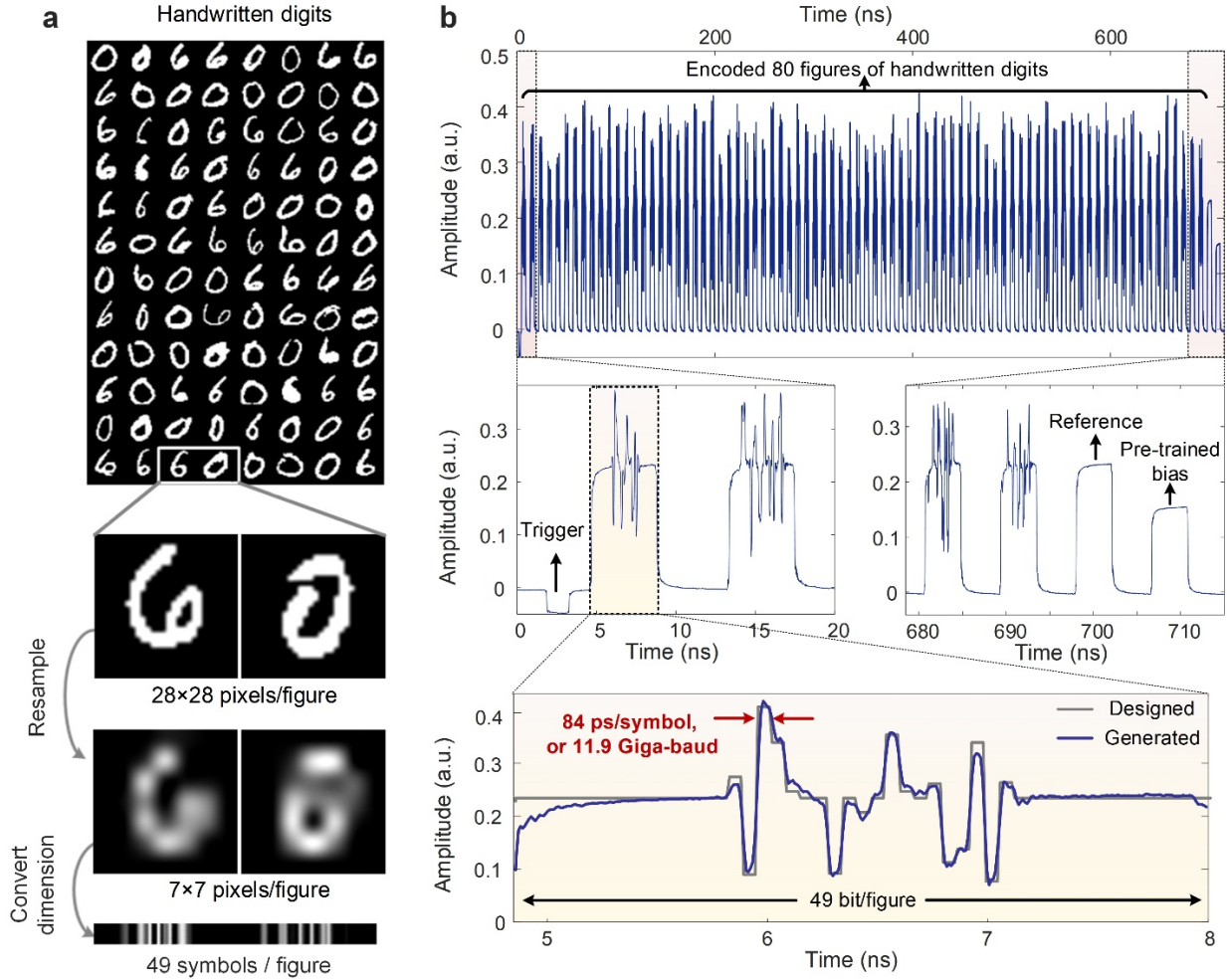

Fig. 5. Time-domain multiplexed input layer of handwritten digits 0 and 6. a, Preprocessing flow of the handwritten digits test. Each handwritten digit figure was a 28×28 array of gray-scale pixels. To match the number of input nodes—49 in our case, the figures were resampled to 7×7 pixels. Then the gray scale data was rearranged into a one-dimensional array. Negative neuron connections were achieved by multiplying the data stream with the symbols of pre-trained weights. b, Generated 11.9 Giga-baud data stream for the encoded 80 figures of the handwritten digits, showing 49-symbol encoded data for each figure and 3 symbols padded for post measurement, including a trigger symbol to trigger the oscilloscope, a reference symbol to calibrate the reference level, and a bias symbol encoded with the pre-trained bias to locate the decision boundary.

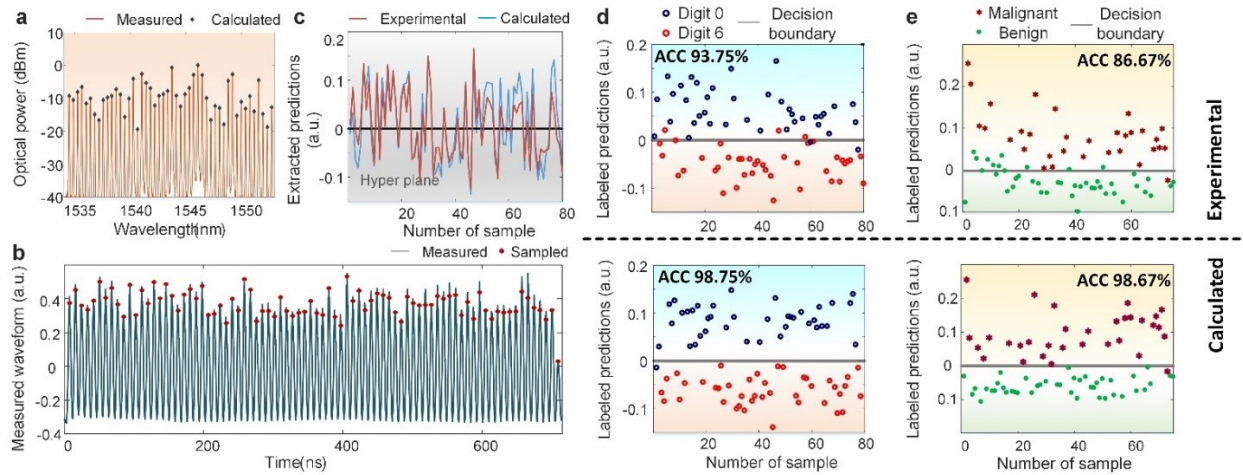

Fig. 6. Experimental classification of the handwritten digits and cancer cells. a, Optical spectrum of the shaped soliton crystal micro-comb for the handwritten digits recognition. b, Measured and sampled output waveform from the photodetector. c, Recovered ONN predictions X×W+b of the handwritten digits acquired by rescaling the sampled results via the reference symbol, and the decision boundary given by the hyper-plane X×W+b=0 (black line). d, Predictions for handwritten digit recognition labeled according to their correct answers, showing an accuracy (ACC) of 93.75% ("Experimental"). This is compared to the numerical results, calculated offline using the designed ONN parameters, which had an accuracy of 98.75% ("Calculated"). e, Labelled predictions for benign versus malignant tumor cell classification, showing an accuracy of 86.67% for the ONN, and 98.67% for the numerical results.

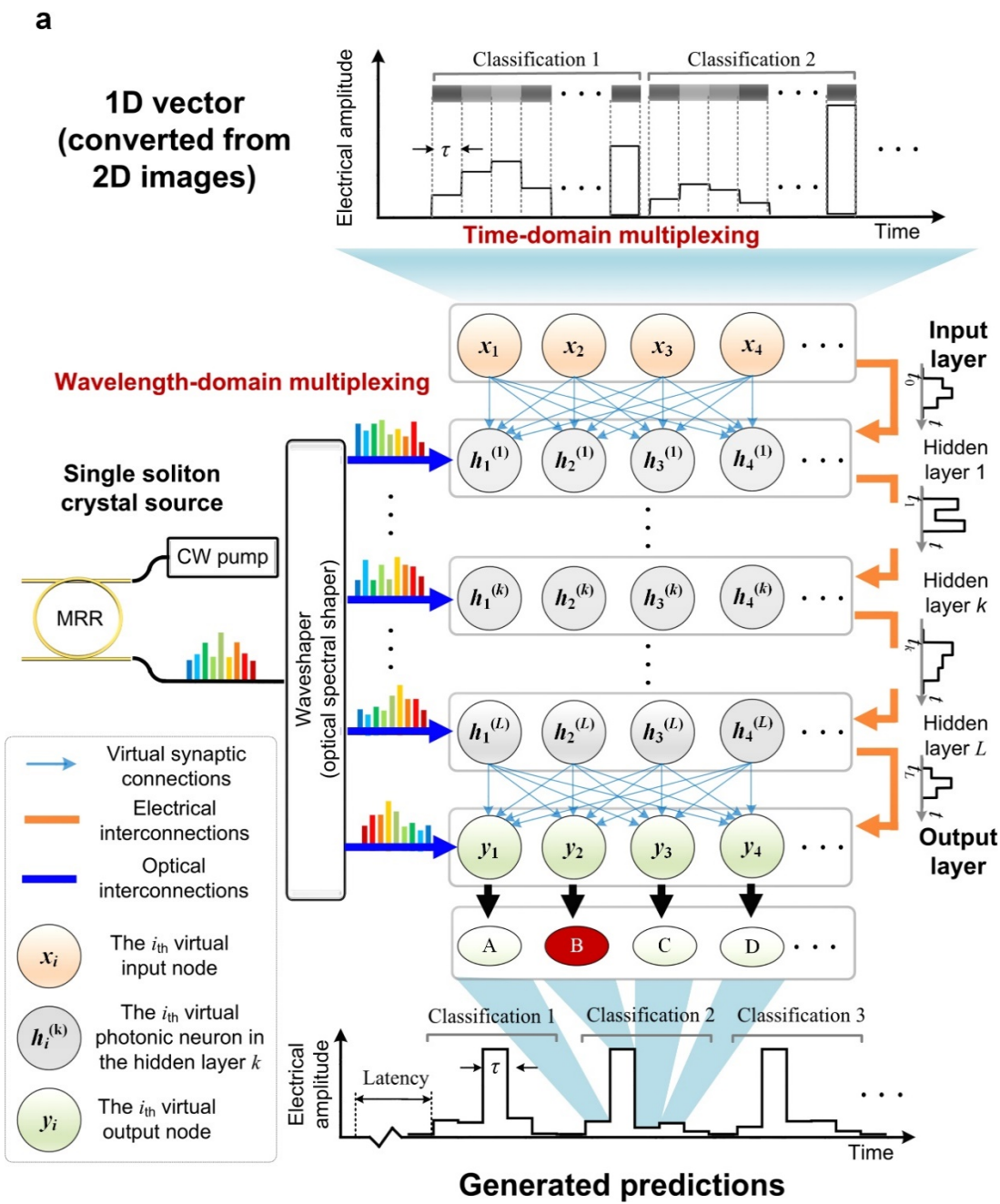

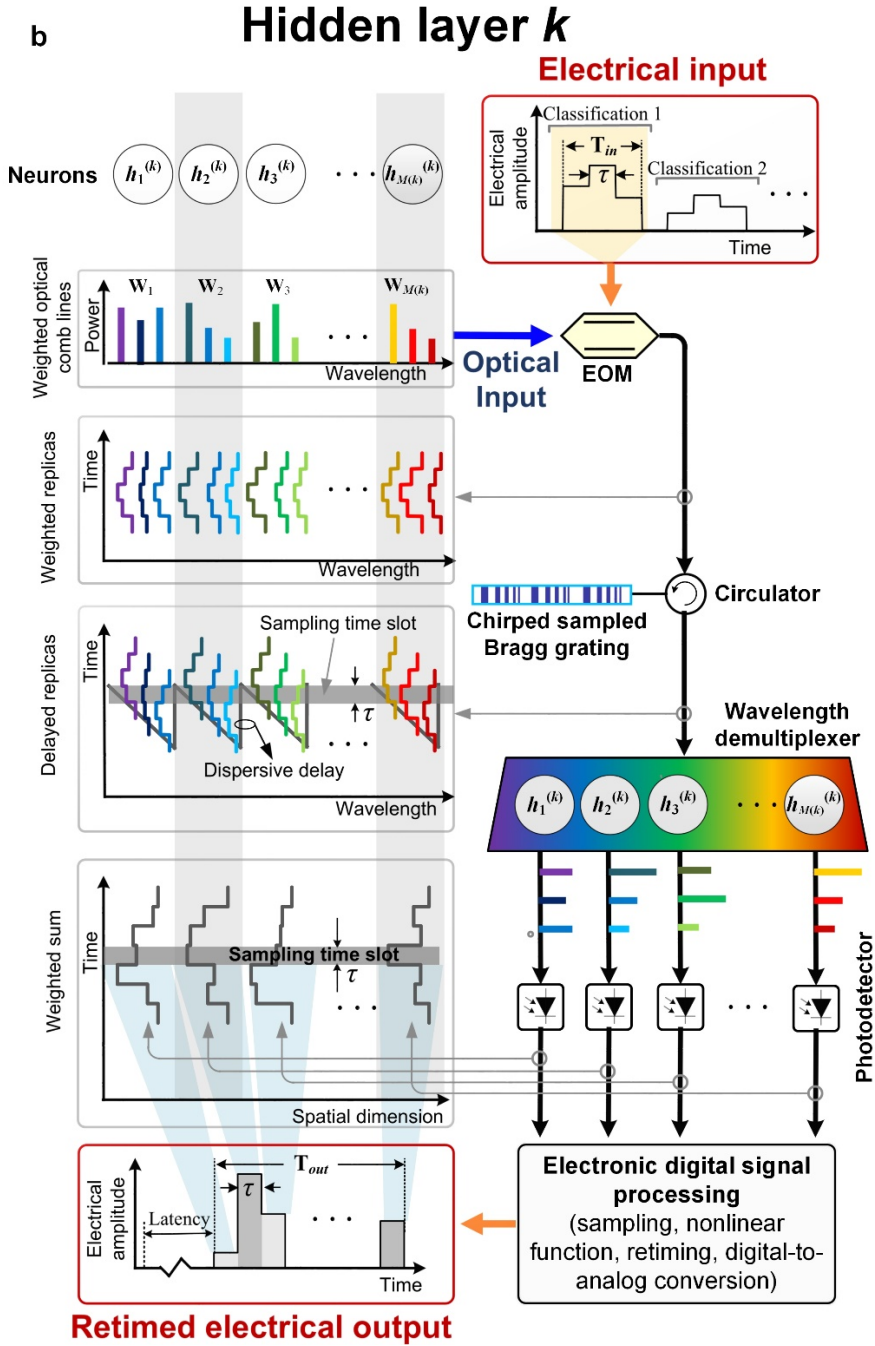

Fig. 7. Designed deep optical neural network based on micro-combs. a, Schematic of the full multilayer ONN. The shaded region indicates the scaled part of the designed deep neural network. The full network is composed by the input layer, $L$ hidden layers ($L$=1, 2, 3, …) with the $k$th layer containing $M(k)$ neurons ($M(k)$ is an integer), and an output layer that is constituted by $M(L+1)$ neurons. The raw input data stream contains multiple equal-size 2D data samples, each is first converted into a 1D vector with a length of $N$ and then sequentially multiplexed into a temporal waveform via electrical digital-to-analog conversion. b, Detailed schematic of layer $k$, illustrating how multiple neurons within each layer are implemented.